\pdfoutput=1

\documentclass[aps,prb,reprint,superscriptaddress]{revtex4-1}

\usepackage[utf8]{inputenc} 

\usepackage{graphicx} 
\usepackage{epstopdf}
\usepackage{siunitx}
\usepackage{hyperref}

\usepackage{booktabs} 
\usepackage{array} 
\usepackage{paralist} 
\usepackage{verbatim} 
\usepackage[nolist]{acronym} 
\usepackage{amsmath}
\usepackage{mathrsfs}
\usepackage{bm}
\usepackage[british]{babel}

\usepackage{hyphenat}

\begin{document}

\title{High resolution spectroscopy of individual erbium ions in strong magnetic fields.}
\author{Gabriele G. \surname{de Boo}}
\email{g.deboo@unsw.edu.au}
\affiliation{Centre for Quantum Computation and Communication Technologies, UNSW Sydney, Australia}
\author{Chunming \surname{Yin}}
\email{c.yin@unsw.edu.au}
\affiliation{Centre for Quantum Computation and Communication Technologies, UNSW Sydney, Australia}
\author{Milo\v s \surname{Ran\v ci\'c}}
\affiliation{Centre for Quantum Computation and Communication Technologies, The Australian National University, Australia}
\affiliation{Quantronics group, Service de Physique de l'Etat Condens, CEA Saclay, France}
\author{Brett C. \surname{Johnson}}
\affiliation{Centre for Quantum Computation and Communication Technologies, University of Melbourne, Australia}
\author{Jeffrey C. \surname{McCallum}}
\affiliation{Centre for Quantum Computation and Communication Technologies, University of Melbourne, Australia}
\author{Matthew \surname{Sellars}}
\affiliation{Centre for Quantum Computation and Communication Technologies, The Australian National University, Australia}
\author{Sven \surname{Rogge}}
\affiliation{Centre for Quantum Computation and Communication Technologies, UNSW Sydney, Australia}

\begin{abstract}
In this paper we use electrically detected optical excitation spectroscopy of individual erbium ions in silicon to determine their optical and paramagnetic properties simultaneously. We demonstrate that this high spectral resolution technique can be exploited to observe interactions typically unresolvable in silicon using conventional spectroscopy techniques due to inhomogeneous broadening. In particular, we resolve the Zeeman splitting of the \textsuperscript{4}I\textsubscript{15/2} ground and \textsuperscript{4}I\textsubscript{13/2} excited state separately and in strong magnetic fields we observe the anti-crossings between Zeeman components of different crystal field levels. We discuss the use of this electronic detection technique in identifying the symmetry and structure of erbium sites in silicon.
\end{abstract}

\maketitle
\date{\today}

\begin{acronym}
	\acro{EPR}[EPR]{electron para\-magnetic resonance}
	\acro{YSO}[Y\textsubscript{2}SiO\textsubscript{5}]{yttrium orthosilicate}
	\acro{ErYSO}[Er:Y\textsubscript{2}SiO\textsubscript{5}]{erbium doped yttrium orthosilicate}
	\acro{4I13/2}[\textsuperscript{4}I\textsubscript{13/2}]{\textsuperscript{4}I\textsubscript{13/2}}
	\acro{4I15/2}[\textsuperscript{4}I\textsubscript{15/2}]{\textsuperscript{4}I\textsubscript{15/2}}
	\acro{EDFA}[EDFA]{erbium doped fiber amplifier}
	\acro{PL}[PL]{photo\-lumi\-nescence}
	\acro{PLE}[PLE]{photo\-lumi\-nescence exci\-ta\-tion}
	\acro{EXAFS}[EXAFS]{extended x-ray absorption fine structure}


\end{acronym}

\section{Introduction}
Erbium doped silicon has been investigated since the 1980s in order to establish optical emission and absorption of silicon at 1550 \si{\nano\metre}\cite{Ennen1983,Tang1989,Kenyon2005}. This wavelength is used for long distance optical communication as this is where optical fibers have their lowest losses. Erbium doping allows glasses and crystals like silicon to fluoresce at 1550 \si{\nano\metre} by using erbium's optical transition between the \acs{4I15/2} and \acs{4I13/2} state. While this has enabled successful technologies such as the \ac{EDFA} used for fiber communication, erbium doped silicon devices are presently not implemented because the fraction of erbium ions that are optically active is low and the \ac{PL} is quenched at room temperature, resulting in low efficiencies. To develop better silicon photonic devices it is necessary to identify which erbium ions in silicon are efficient emitters and what sites they occupy in the crystal.

Erbium has been found to occupy a variety of sites in silicon and \ac{PL} experiments have been used to resolve the site symmetry through the number of crystal field lines in the spectrum~\cite{Tang1989,Przybylinska1996PRB}. Several studies have found evidence of erbium occupying a tetrahedral interstitial site~\cite{Wahl1997,Assali2003,Tang1989,Przybylinska1996PRB}, as well as sites with lower symmetry~\cite{Przybylinska1996PRB,Demaatgersdorf1995,Vinh2003}. While \acs{PL} studies are able to resolve the crystal field splitting, they provide limited knowledge of the microscopic structure of the defect centers. Conversely, Electron paramagnetic resonance (\acs{EPR}) studies can provide more information about the microscopic structure of paramagnetic ions by examining the magnetic field direction dependence of the Zeeman splitting of the ground state doublet~\cite{Abragam1970}. For erbium in silicon, the \ac{EPR} signal of several low symmetry sites has been identified~\cite{Carey1999PRB}. Because the symmetry is low, the \ac{EPR} results do not provide enough information about the crystal field to associate these sites with the crystal field splitting of sites found in \acs{PL} experiments.

Since the optical transitions of the \acs{PL} lines involve the same spin states accessible to \acs{EPR} experiments, observing the splitting of the \acs{PL} lines in a magnetic field could match sites between both types of experiments. In magneto-optical experiments, the Zeeman splitting of the optical transitions results from the splitting both in the \acs{4I15/2} state and the \acs{4I13/2} state. The Zeeman splitting depends on the site symmetry and the magnetic field direction and if measured for multiple crystal field levels, a detailed model of the crystal field structure can be created~\cite{Dieke1968}. It also makes it possible to identify ground states with no transverse Zeeman splitting, which is not possible with \acs{EPR}.

However, for the \acs{PL} lines of erbium in silicon, the inhomogeneous broadening generally is such that the lines do not split in a magnetic field, but broaden and dissolve~\cite{Vinh2004}. Only if a \acs{PL} line results from a single site and its inhomogeneous broadening is smaller than the Zeeman splitting, is it possible to deduce its magneto-optical properties~\cite{Thonke1988,Sun2008}. This has been achieved for one particular erbium site in silicon where the inhomogeneous broadening is as small as \num{2} \si{\giga\hertz}~\cite{Vinh2003,Vinh2004}. Although the site symmetry could be determined for this site from the Zeeman splitting, it does not correspond to the symmetry of any of the sites that have been found in other optical or \acs{EPR} experiments of erbium in silicon. Therefore, it remains necessary to characterize more sites. 

In this study we overcome the inhomogeneous broadening by studying the optical transitions of individual erbium ions. In a previous study, we demonstrated that the optical absorption of erbium ions in silicon can be detected using a charge sensing process~\cite{Yin2013}. This method is based on optical excitation and allows us to measure the spectrum of individual erbium ions in arbitrary magnetic fields, whereas \acs{EPR} studies are limited to a small range of magnetic fields by the cavity resonance condition. Here, we use this method to perform high resolution spectroscopy and we resolve the Zeeman splitting of both the ground state and the excited state, providing a link between optical spectroscopy and \acs{EPR} spectroscopy. We also use this technique in large magnetic fields, where we observe non-linear Zeeman interactions due to interactions with other crystal field levels, as well as an avoided crossing between crystal field levels.

\section{Experiment}
For this study, we have used a \textsuperscript{167}Er (I=7/2) implanted fin field-effect transistor (finFET) with a channel width and length of \num{890} \si{\nano\metre} and \num{80} \si{\nano\metre}, respectively. We estimate that these finFETs contain 30-40 Er atoms that contribute to the measured signal. After ion implantation the finFETs were annealed at 800\si{\degreeCelsius} to optically activate the Er without causing degradation to the device characteristics. 

We illuminate the device with a tunable laser to excite the Er optical transitions from the \acs{4I15/2} state to the \acs{4I13/2} state, which are approximately \num{800} \si{\milli\electronvolt} apart. We use a wavemeter to monitor the wavelength of the laser in order to increase the accuracy of determining the transition wavelength~\cite{Zhang2019}. We find more than 20 groups of erbium resonances over a wavelength range between \num{1529} and \num{1546} \si{\nano\metre}. The resonances in each group are spread over frequency spans between 0.3 and 5 \si{\giga\hertz}, with up to 20 individual resonances per group. Their line width is typically \num{100} \si{\mega\hertz} or less. We identify these resonances as the hyperfine splitting of the optical transitions of individual erbium ions~\cite{Yin2013}. The groups of resonances can originate from different ions or several groups of resonances can originate from the same ion if they are transitions to different crystal field levels of that ion. Due to variations in site symmetry and local environment, different erbium ions have optical transitions at different wavelengths, which makes it possible for us to distinguish between them.

In order to investigate the paramagnetic properties of the erbium ions we apply a magnetic field and observe how the Zeeman effect splits the optical transitions. How the optical transitions split and shift due to the magnetic field reflects the splitting of the levels in the \acs{4I15/2} and \acs{4I13/2} states. The \acs{4I15/2} and \acs{4I13/2} states each consist of 16 and 14 states, respectively, which are split by the crystal field into several multiplets spaced hundreds of \si{\giga\hertz} apart~\cite{Kenyon2005,Przybylinska1995}. In sites of cubic symmetry there will be several quadruplets~\cite{Lea1962}, while in low symmetry sites there will only be doublets. These doublets, which can only be split by a magnetic field, are known as Kramers doublets~\cite{Abragam1970}. It can be convenient to describe the linear Zeeman splitting of a Kramers doublet with an effective spin Hamiltonian, where the system is approximated as a spin-1/2 particle with a certain g factor~\cite{Abragam1970}. The energy splitting of a Kramers doublet in a magnetic field is then given by $\mathbf{E}=g\mu_{B}\mathbf{B}$.

In this work, experiments were carried out at 4.2 \si{\kelvin} and only the lowest crystal field level of the \acs{4I15/2} was populated. All optical excitation is due to the excitation of this ground state. We have not identified any quadruplets and we therefore assume that the Er exists on low symmetry sites. 

Previously we observed that the optical transitions split in a magnetic field into two Zeeman branches~\cite{Yin2013}. Between two Kramers doublets in the \acs{4I15/2} and \acs{4I13/2} states there are four possible energy differences. If only two optical transitions are observed, they either represent the difference or the sum in Zeeman splitting in both doublets. The difference in transition frequency $\Delta\nu$ between the branches is then described by $\Delta \nu = \Delta g \mu B$, where $\Delta g$ represents either the difference or the sum of the splitting of the Kramers doublets in the \acs{4I15/2} and \acs{4I13/2} states. 


\section{Quadratic Zeeman}

\begin{figure}
    \includegraphics[width=\columnwidth]{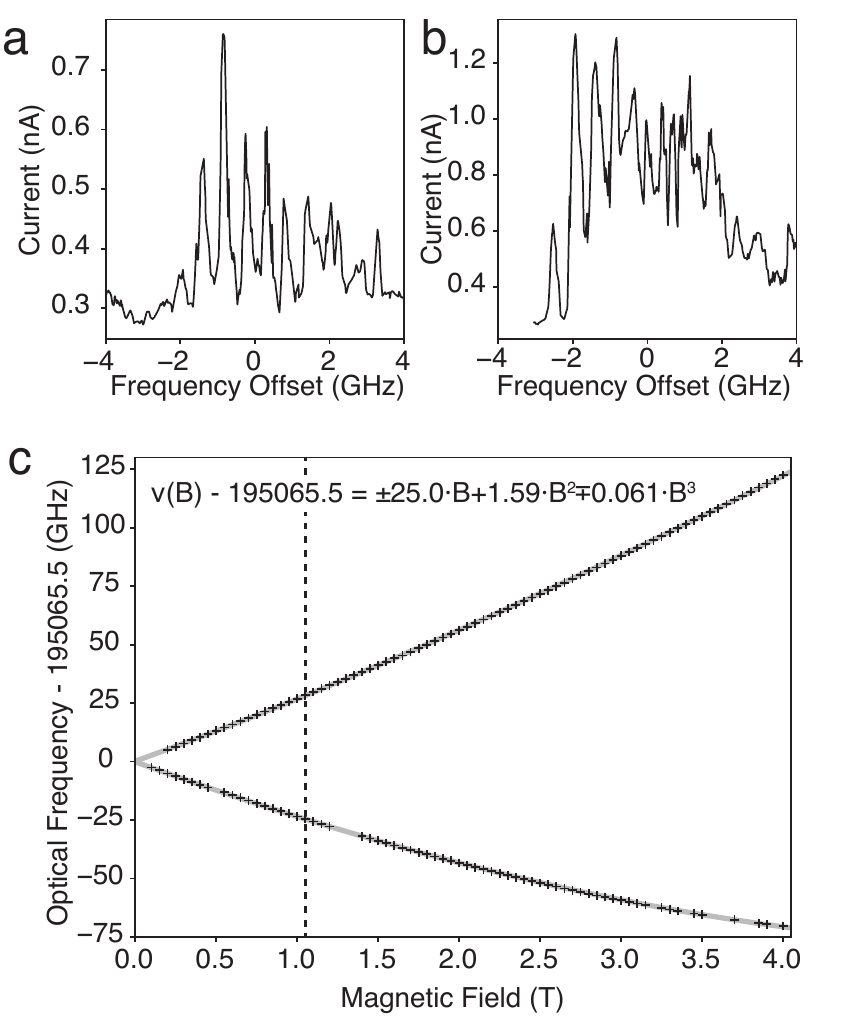}
    \caption{\label{fig1}Hyperfine spectra of the two Zeeman branches of a single erbium ion at 1.05 \si{\tesla} (a and b). The location of the hyperfine spectra as a function of magnetic field is shown in (c) together with a model that describes the Zeeman splitting. The dotted line indicates the magnetic field where (a) and (b) were measured.}
\end{figure}

We measure the excitation spectra of several groups of resonances as a function of magnetic field in order to examine their Zeeman splitting. Fig. \ref{fig1}a and \ref{fig1}b show spectra measured at \num{1.05} \si{\tesla} in the low and high frequency Zeeman branches, respectively, of a group of resonances at \num{195065.5} \si{\giga\hertz} (\num{1536.881} \si{\nano\metre}). The spectra consist of a multitude of resonances spread over 5 \si{\giga\hertz} that are associated with different hyperfine transitions. At magnetic fields below 0.2 \si{\tesla} the electronic Zeeman interaction and the hyperfine interaction are of similar strength and the Zeeman branches can not be resolved separately. Above 0.2 \si{\tesla} the hyperfine interaction is smaller than the electronic Zeeman and the two Zeeman branches are resolved. In high magnetic fields, the hyperfine interaction can be considered a perturbation to the electronic Zeeman interaction with its size being proportional to the electronic and nuclear magnetic moment in both states. The magnetic moments do not change significantly while the Zeeman splitting is smaller than the crystal field splitting and as a result the hyperfine splitting in this measurement is constant between 0.2 \si{\tesla} and 4.0 \si{\tesla}.  

Using the fact that the hyperfine spectrum does not change above 0.2 \si{\tesla}, we can accurately determine the location of the Zeeman branches by fitting a model of the spectrum to the measured spectrum at each field step. The result is shown in Fig. \ref{fig1}c. We observe a deviation from the linear Zeeman effect which increases as the field is increased. 

The observed non-linearity in the Zeeman effect can be understood from perturbation theory. When the Zeeman splitting of the crystal field levels is much smaller than the splitting between the crystal field levels, the Zeeman effect can be considered a small perturbation and it can be described by perturbation theory. The first order perturbation leads to a linear Zeeman splitting that is proportional to the magnetic moment of the two states in the doublet. The g-factor in the spin-1/2 description of a Kramers doublet represents this linear Zeeman splitting. The second and third order perturbations result from the interactions between crystal field levels and lead to quadratic and cubic Zeeman shifts.

The spin-1/2 description used previously ignores higher order perturbation terms. This is valid for \acs{EPR} experiments because the quadratic term from the second order perturbation is the same for both levels and cancels out in the transition between the levels, while the third order component is too small to be significant at the magnetic fields used in \acs{EPR} experiments. In optical excitation experiments however, the quadratic Zeeman does not cancel out because the energy difference is measured between a Kramers doublet in the \acs{4I15/2} state and a Kramers doublet in the \acs{4I13/2} state, which in general will not both have the same quadratic Zeeman effect. Consequently, non-linear Zeeman effects have been observed for the optical transitions of erbium in various hosts, including silicon~\cite{Wannemacher1989,Vinh2003,Bottger2006,Marino2016}.

Taking this into account, we can describe the non-linearity in the optical transitions in Fig. \ref{fig1}c as the difference in quadratic and cubic Zeeman splitting between the \acs{4I15/2} and \acs{4I13/2} states. We fit the transition frequencies in Fig. \ref{fig1}c as a function of magnetic field in both branches to a third order polynomial where the linear and cubic terms ($\Delta g$ and $\Delta L$) are the same size, but opposite sign between the two branches, while the quadratic term ($\Delta K$) is the same size and sign for both branches~\cite{Macfarlane1984,vonLindenfels2013}:
\begin{equation}
\nu(B) = \nu_{0} \pm \Delta g \mu_{B} B + \Delta KB^{2} \pm \Delta LB^{3}
\end{equation}
The fit results in a $\Delta g$ of \num{1.79 \pm 0.02} (\num{25.0 \pm 0.3} \si{\giga\hertz\per\tesla}), a $\Delta K$ of \num{1.59 \pm 0.04} \si{\giga\hertz\per\tesla\squared} and a $\Delta L$ of \num{-0.061 \pm 0.003} \si{\giga\hertz\per\tesla\cubed}. The value for $\Delta K$ is comparable to the quadratic Zeeman values reported in references \cite{Vinh2003} and \cite{Wannemacher1989}. Although the cubic Zeeman term is small, its influence is not insignificant at high fields and by including it the uncertainty in the linear term is reduced.

While the quadratic Zeeman shift of erbium's optical transitions is rarely resolved in experiments, the linear Zeeman splitting is commonly used to determine the symmetry and crystal field structure of erbium~\cite{Dieke1968,Ammerlaan2001}. When the the linear Zeeman splitting is combined with the energy splitting between crystal field levels, the parameters of a crystal field Hamiltonian that describes all the levels in a total angular momentum manifold, e.g. \acs{4I15/2} or \acs{4I13/2}, can be estimated~\cite{Przybylinska1996PRB}. Using second order perturbation theory, the quadratic Zeeman component can be calculated from such a Hamiltonian and its value can be compared to the observed quadratic Zeeman shift. This extends the number of observations that contribute to estimating the crystal field parameters, thereby improving the accuracy of the model.

A limitation to the accuracy of determining the Zeeman splitting in this experiment is the accuracy of the magnetic field strength due to misalignment and hysteresis in the superconducting magnet. For this experiment we used a 12 \si{\tesla} superconducting magnet with a inductance of \num{20.36} \si{\henry}. We estimate that the uncertainty due to misalignment of the device with the center of the magnet can cause the field in the device to be up to 1\% less than the set value. We found with a cryogenic Hall sensor that the hysteresis of the magnet can be as large as 10 \si{\milli\tesla}. These uncertainties have been taken into account into determining the uncertainties for $\Delta g$, $\Delta K$ and $\Delta L$.

\section{Determination of the g factor}
\begin{figure}
    \includegraphics[width=\columnwidth]{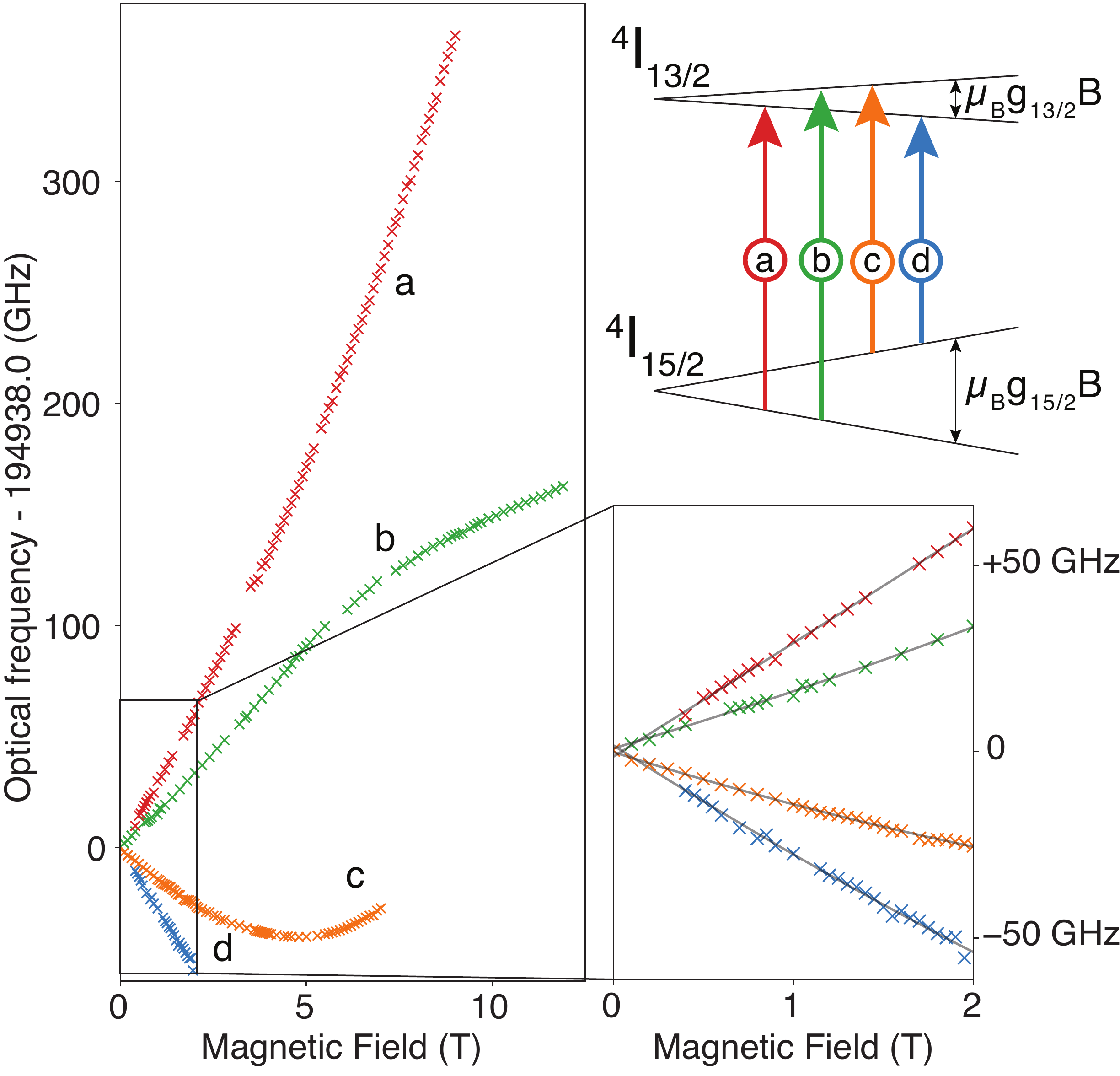}
    \caption{\label{like-unlike}Zeeman splitting of an erbium transition at 194938 \si{\giga\hertz} (1537.886 \si{\nano\metre}). All four optical transitions between the two Kramers doublets are visible below 2 \si{\tesla}. The insets show the corresponding energy level diagram as B increases and a zoom-in of the low field region.}
\end{figure}

Unlike the Zeeman splitting of the resonance observed in Fig.~\ref{fig1}, for several erbium resonances we observe all four optical transitions between the two Kramers doublets of the ground and excited states. An example of this is shown in figure~\ref{like-unlike}, where the peak locations of an erbium resonance at \num{194938.0} \si{\giga\hertz} (\num{1537.886} \si{\nano\metre}) are shown. The four transitions are labeled following Ref.~\cite{Bottger2006}. When all four transitions are visible, it is possible to determine the Zeeman splitting of both Kramers doublets separately. Whether transition \emph{b} or \emph{c} has a larger frequency depends on whether the Zeeman splitting is larger in the \acs{4I15/2} or \acs{4I13/2} state. The transitions are identified by observing which ones lose strength and disappear at high field due to the thermal depopulation of the upper level in the \acs{4I15/2} doublet. For the transitions shown in figure \ref{like-unlike}, transition \emph{c} vanishes above \num{7} \si{\tesla}, from which we conclude that it originates from the upper level and consequently transition \emph{b} originates from the lower level. Transition \emph{d} also originates from the upper \acs{4I15/2} level, but due to a lower signal strength the thermal depopulation causes it to vanish at a lower field (2 \si{\tesla}). 

\begin{figure}
	\includegraphics[width=\columnwidth]{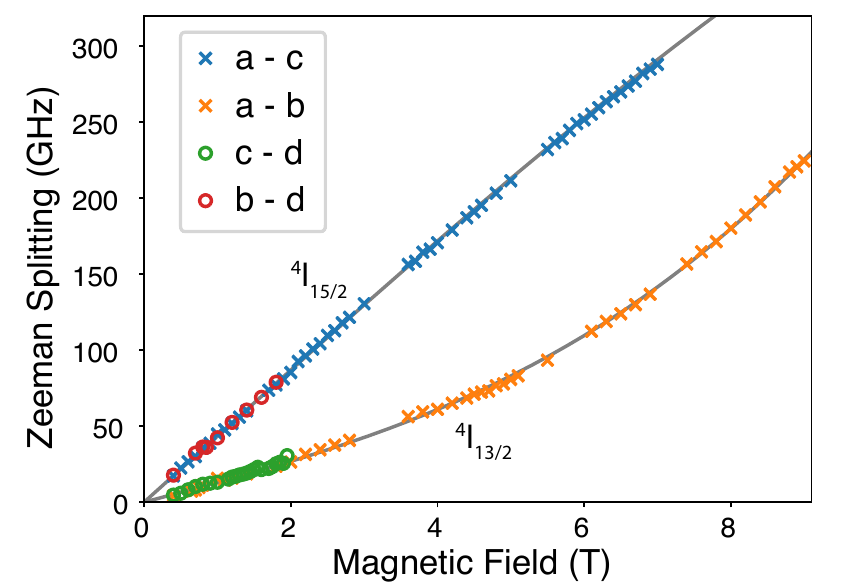}
	\caption{\label{field_dependent_zeeman}Zeeman splitting as a function of magnetic field, determined from the frequency difference between optical transitions \emph{a}, \emph{b}, \emph{c} and \emph{d} in Fig. \ref{like-unlike}. For each state, the linear and cubic Zeeman splitting have been fit (solid lines).}
\end{figure}

For magnetic fields larger than 2 \si{\tesla} the Zeeman splitting in Fig. \ref{like-unlike} becomes non-linear due to higher order Zeeman effects, similar to the observation in Fig. \ref{fig1}. Fig. \ref{field_dependent_zeeman} shows the Zeeman splitting of the \acs{4I15/2} and \acs{4I13/2} states separately, as determined from the frequency differences between the transition frequencies. The largest frequency difference that could be determined for the \acs{4I15/2} state before the optical transitions from the upper level vanish, is \num{288} \si{\giga\hertz} or \num{3.3}$\mathrm{\times kT}$ (T = \num{4.2} \si{\kelvin}, kT = \num{87.5} \si{\giga\hertz}). Fig. \ref{field_dependent_zeeman} shows that both states have linear and non-linear Zeeman spitting, with the splitting in the \acs{4I15/2} being larger than in the \acs{4I13/2} state. The non-linearity at high fields can not be caused by the quadratic Zeeman effect, because it is the same for both levels and cancels out. The splitting in each state was fit to 
\begin{equation}
	f(B) = g \mu_{B} B + L B^{3}
\end{equation}
where $g$ and $L$ describe the linear and cubic Zeeman splitting, respectively. For the \acs{4I15/2} state we obtain $g=\num{3.12 \pm 0.05}$ and $L$=\num{-0.049 \pm 0.005} \si{\giga\hertz\per\tesla\cubed} and for the \acs{4I13/2} state we obtain $g$=\num{0.92 \pm 0.01} and $L$=\num{0.151 \pm 0.005} \si{\giga\hertz\per\tesla\cubed}. From this splitting can be concluded that the site is not cubic because the g is either 6 or 6.8 for the ground state doublet of a cubic site~\cite{Abragam1970}. This implies that the erbium ion occupies a site of lower symmetry and from this follows that the Zeeman splitting should be anisotropic.

Whether two or four lines are visible for an erbium transition is determined by the transition rules and the precise composition of the wave function. The erbium crystal field levels are superpositions of total angular momentum projection ($m_{J}$) states, where $J \in -15/2:+15/2$ for the \acs{4I15/2} state and $J \in -13/2:+13/2$ for the \acs{4I13/2} state. In general, if the mixture is such that for a pair of transitions (\emph{a} and \emph{d} or \emph{b} and \emph{c} in Fig. \ref{like-unlike}) the wave function has to change by more than $m_{J}=\pm1$, the transition is forbidden. For some mixtures only two transitions are allowed and for others all four. To determine the precise composition of the wave function would require knowledge of the site symmetry and identification of the particular crystal field levels. This involves measuring the Zeeman splitting along arbitrary directions for most crystal field levels in both the \acs{4I15/2} and the \acs{4I13/2} manifolds. The experiment discussed in this paper examines the Zeeman splitting of one crystal field level in one field direction, which does not provide enough information do to this. 

\section{Avoided Crossing}
\begin{figure}
	\includegraphics[width=\columnwidth]{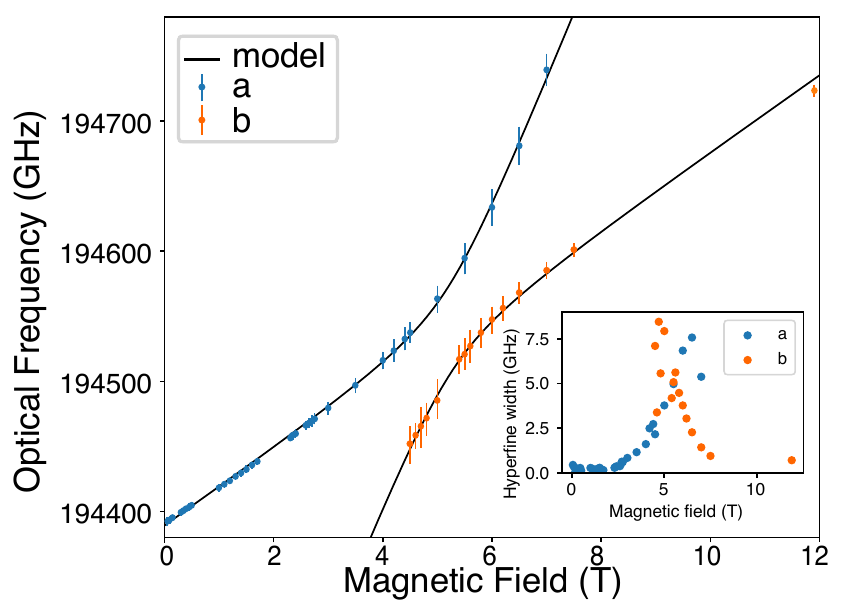}
	\caption{\label{avoided_crossing}Avoided crossing between two crystal field levels within the \acs{4I13/2} manifold. At the avoided crossing the width of the hyperfine splitting changes in both transitions, which is indicated with vertical lines (shown times 10 for emphasis) and in the inset.}
\end{figure}

When the Zeeman splitting of crystal field levels becomes comparable to the splitting between them, the interaction between the crystal field levels can result in avoided crossings between them~\cite{Crosswhite1961}. Figure \ref{avoided_crossing} shows a resonance at \num{194391} \si{\giga\hertz} at \num{0} \si{\tesla} that has an avoided crossing with a higher lying crystal field level at \num{5.3} \si{\tesla}. The transition strength and the hyperfine spectrum change at the magnetic field where the two lines have an avoided crossing. The line starting at \num{194391} \si{\giga\hertz} has a total hyperfine width of 0.3 \si{\giga\hertz} at low magnetic field. As it nears the avoided crossing, the hyperfine width increases and the resonances become weaker until they disappear above \num{6} \si{\tesla}. The second line starts out weak and wide below 6 \si{\tesla} and becomes narrower and stronger towards higher magnetic fields.  

Without a magnetic field the splitting between the two levels in the avoided crossing is approximately \num{400} \si{\giga\hertz}, significantly larger than the thermal energy. As the higher lying level is visible at low fields it can not originate from an excited level of the \textsuperscript{4}I\textsubscript{15/2} state as it would not be populated. The avoided crossing must therefore be occurring within the \textsuperscript{4}I\textsubscript{13/2} manifold. 

The inset in Fig. \ref{avoided_crossing} shows the width of the hyperfine spectrum for both lines as a function of magnetic field. The total hyperfine spreading of the optical transition is determined either by the difference in hyperfine splitting between the \acs{4I15/2} and the \acs{4I13/2} states or the sum, depending on whether the nuclear spin states have the same order or not. If the two crossing crystal field levels have the opposite order of hyperfine levels, the hyperfine splitting of the optical transition will transfer from being the difference to the sum or vice versa along the avoided crossing. This results in a change in the hyperfine spreading such as observed in the inset of Fig. \ref{avoided_crossing}.


The measured resonance locations have been fit to a two level Hamiltonian in order to give an indication of the interaction strength between the two levels. This Hamiltonian is defined as
\begin{equation}
    \mathrm{H} = \left( \begin{array}{cc}
    \mathrm{E_{1}} + \mathrm{\mu_{B}} \mathrm{g_{eff,1}}\mathrm{B}   & \mathrm{W} \\
    \mathrm{W}                   & \mathrm{E_{2}} + \mathrm{\mu_{B}} \mathrm{g_{eff,2}}\mathrm{B} \\  
    \end{array} \right)
\end{equation}
where $E_{1}$ and $E_{2}$ are the transition energies without a magnetic field, $g_{eff,1}$ and $g_{eff,2}$ the effective g factors of the optical lines and $W$ the interaction strength between the two levels. The fitting result is shown as a solid line in Fig. \ref{avoided_crossing}. The interaction strength resulting from this fit is \num{35} \si{\giga\hertz} or \num{180} \si{\micro\electronvolt}. The model estimates a splitting between the levels of \num{392} \si{\giga\hertz} at \num{0} \si{\tesla}. In general, the interaction strength at the avoided crossing will depend on the magnetic field direction as a magnetic field component perpendicular to the symmetry axis will mix different terms in the Hamiltonian than a parallel field would do~\cite{Crosswhite1961}. Measurements of an avoided crossing along different magnetic field directions would give insight into the crystal field Hamiltonian, but this was beyond the scope of this experiment as this would require the ability to rotate large magnetic fields.


\section{Conclusion}
Our results demonstrate that optical excitation spectroscopy of individual erbium ions in silicon can be used to measure their paramagnetic properties. From the splitting of the optical transitions in a magnetic field, the Zeeman splitting of both the \acs{4I15/2} ground state and the \acs{4I13/2} excited state can be determined in a range of magnetic fields. Because of the narrow line width of individual ions this is possible at magnetic fields well below 1 \si{\tesla}. If an isotope without nuclear spin is used, the required field strength will be even smaller. In contrast, measurement of the Zeeman splitting in ensembles requires large magnetic fields to overcome inhomogeneous broadening~\cite{Vinh2004}.  

Further understanding of erbium's level structure requires the site symmetry to be known. These results show that it should be possible to measure the g tensors in the \acs{4I15/2} and \acs{4I13/2} states of individual erbium ions by varying the magnetic field direction. These g tensors can then be used together with the interactions between crystal field levels to determine the site symmetry and crystal field structure.

\begin{acknowledgments}
We acknowledge the AFAiiR node of the NCRIS Heavy Ion Capability for access to ion-implantation facilities. C.Y. acknowledges support from an ARC Discovery Early Career Researcher Award (Grant DE150100791). This work was supported by the ARC Centre of Excellence for Quantum Computation and Communication Technology (Grant CE170100012) and the Discovery Project (Grant DP150103699).
\end{acknowledgments}

\bibliographystyle{h-physrev}
\bibliography{library}

\end{document}